\documentclass[
  journal=pasa,
  manuscript=research-paper, 
  year=2025,
  volume=YY,
]{cup-journal}

\usepackage{amsmath}
\usepackage{amssymb,microtype,siunitx,booktabs}
\usepackage{mathrsfs}
\usepackage[breaklinks,colorlinks,
   urlcolor=blue,citecolor=blue,linkcolor=blue]{hyperref}
\newcommand{\fig}{Figure}

\newcommand{\sect}{Section}

\newcommand{\eqn}{Equation}

\title{The effect of interstellar scattering on coherent radio emission from stars: the case of CU Vir}

\author{J.~S.~Morgan}
\affiliation{CSIRO, Space and Astronomy, P.O. Box 1130, Bentley, WA 6102, Australia}
\email[J.~S.~Morgan]{john.morgan@csiro.au}

\author{B. Das}
\affiliation{CSIRO, Space and Astronomy, P.O. Box 1130, Bentley, WA 6102, Australia}
\alsoaffiliation{National Centre for Radio Astrophysics, Tata Institute of Fundamental Research, Pune University Campus, Pune-411007, India}

\author{H.~E.~Bignall}
\affiliation{Manly Astrophysics, 15/41-42 East Esplanade, Manly, NSW 2095, Australia}
\alsoaffiliation{Visiting Scientist, CSIRO, Space and Astronomy, P.O. Box 1130, Bentley, WA 6102, Australia}

\received {dd Mmm YYYY}
\revised  {dd Mmm YYYY}
\accepted {dd Mmm YYYY}
\published{22 September 202X}


\begin{document}

\begin{abstract}
A subset of magnetic stars exhibit periodic radio pulses produced by the coherent electron cyclotron maser mechanism.
These pulses are known to exhibit both temporal and spectral variations, which have been attributed to phenomena intrinsic to the stellar magnetosphere.
However, in order to fully characterise the radio pulses and use them as magnetospheric probes (as suggested by past studies), it is also important to  consider the effects of phenomena extrinsic to the magnetosphere.
In this paper, we investigate whether interstellar scintillation could be a relevant mechanism for explaining spectral and temporal variations observed for coherent stellar radio emission.
For that, we consider the case of the well-characterised magnetic hot star CU\,Vir.
At 400 MHz, coherent radio emission from the star was reported to exhibit a peculiar spectral evolution that remains unexplained.  
We show that a plausible level of turbulence along the line of sight can produce the observed phenomenon of spectral features.
Our analysis shows that diffractive interstellar scintillation can have a strong effect on the observed dynamic spectrum of radio emission from stars, for an assumed size of the emitting region of $0.01r_\odot$, and that caution should therefore be taken in separating intrinsic and extrinsic features, particularly at low frequencies.
These results are preliminary and further work is required to fully model the scintillation of ECME from stars (in particular the change in source location with frequency), and to explore the full range of plausible scintillation parameters.
We suggest how further observations may be used to test the interstellar scintillation hypothesis.
\end{abstract}

\section{INTRODUCTION}
\label{sec:introduction}

CU\,Vir is one of the most well-studied radio stars owing to the fact that it is the first main-sequence star discovered to emit periodic radio pulses similar to the pulsars \citep{trigilio2000}. It is a chemically peculiar late B star known to harbour large-scale, stable dipolar magnetic field \citep[e.g.][]{kochukhov2014}. Considering the similarity of the observable signature of the pulsed radio emission with that from pulsars, this star was dubbed as a main-sequence pulsar at the time of its discovery. Today, a number of magnetic hot stars are known to exhibit this phenomenon, and they are known as Main-sequence Radio Pulse emitters \citep[MRPs,][]{2021ApJ...921....9D}.

The emission mechanism behind the radio pulses from MRPs is Electron Cyclotron Maser Emission (ECME), which is a coherent phenomenon. It is a beamed emission and the frequency of the emission is proportional to the magnetic field strength at the emission sites.
A detailed review of the salient features of this mechanism is provided in \sect~\ref{subsec:ecme_properties}. 

ECME from CU\,Vir has been detected from around $340$ MHz to $\lesssim 8$ GHz \citep[][etc.]{trigilio2000,leto2006,2021ApJ...921....9D} at more than one epoch. These observations suggest that the observed pulse properties (profile and strength) vary significantly with frequency and also with epoch of observation. In particular, at their lowest frequency of observation ($\approx 400$ MHz), \citet{2021ApJ...921....9D} reported a peculiar spectral feature (see their \fig~11 and our \fig~\ref{fig:dynspec}) where the peak flux density first decreases and then increases with frequency. The origin of this feature remains unclear, although \citet{2021ApJ...921....9D} speculated that it could be due to the superposition of two unrelated bursts occurring at the same time.

Similar to the above peculiar spectral feature, all observed spectral and temporal variations of ECME from MRPs have been attributed to phenomena within their stellar magnetospheres \citep[e.g.][]{trigilio2011,das2020a}. However, one of the consequences of the beamed nature of ECME is that the effective sizes of the emission sites (i.e. the region visible to a terrestrial observer) are $\sim$0.01\,$R_\odot$, much smaller than the size of the star \citep{trigilio2011}. 
Thus, even though MRPs are located relatively close to us ($<400$ pc away), the sites of production of ECME visible to the observer could be extremely compact (explained in \sect~\ref{subsec:ecme_properties}). 
For sufficiently compact emission, another phenomenon comes into play that is known to cause spectral and temporal variation: interstellar scintillation (ISS). While ISS is well established for radio emission from pulsars \citep[which have even more compact emission sites of size$\lesssim$10$^{-3}$\,$R_\odot$; e.g.][]{2012ApJ...758....7G}, its potential role is yet to be explored for stellar bursts. 

The importance of investigating possible roles of phenomena extrinsic to the stellar magnetosphere is that ECME has been shown to be a promising probe for stellar and sub-stellar magnetospheres. In particular, observed spectral variation of ECME has been suggested to carry information regarding the three-dimensional plasma distribution in the stellar magnetosphere \citep{das2020a}.
However, multi-epoch observation of ECME from MRPs is rare. CU\,Vir is the only MRP that has been observed at more than two epochs \citep[e.g.][etc.]{trigilio2000, trigilio2011,2021ApJ...921....9D}, but only at frequencies $\geq 1$ GHz. Consequently, temporal evolution of spectral properties of ECME remains poorly constrained.

In this paper, we make the first attempt to investigate whether scintillation could be relevant for MRPs by taking the case of CU Vir, considering the nature of ECME emission from this star, and determining, via scintillation physics, what impact ISS would have on the dynamic spectrum of the emission if the necessary conditions existed in the interstellar medium along the line of sight.
We also point out recent results suggesting a sufficient density of scattering screens in the local interstellar medium to cause significant scatter along any $\sim70$\,pc line of sight to the Earth.

The paper is organised as follows:
in \sect~\ref{sec:theory} we describe both ISS and the ECME mechanism.
In \sect~\ref{sec:results} we present calculations of dynamic spectra of scintillation based on simulations of scattering screens utilizing the software developed by \citet{2010ApJ...717.1206C} and \citet{2020ApJ...904..104R}.
In \sect~\ref{sec:conclusions} we give our conclusions.

Throughout we assume the following properties for CU Vir:
stellar radius $R_*=2R_\odot$;
polar magnetic field strength $B_\mathrm{p}$=4000\,G;
rotational period $P_\mathrm{rot}$ = $0.5207$\,days;
angle between the line of sight and the stellar rotation axis (inclination angle) $i=46.5^\circ$;
angle between the stellar rotation and magnetic dipole axes (obliquity) $\beta=87^\circ$;
distance = 75.6\,pc \citep{kochukhov2014, shultz2022, gaia2016,gaia2023}.

\section{THEORY AND METHOD}
\label{sec:theory}
\subsection{Interstellar Scintillation (ISS)}
\label{subsec:scintillation}
A simple but sufficient model for most scintillation phenomena is that of a thin, frozen phase screen located between source and observer (\fig~\ref{fig:r_f}).
In the absence of a phase screen, the source would uniformly illuminate a plane at the location of the observer.
The presence of the phase screen means that the plane is instead illuminated by a diffraction pattern.
Temporal variations are induced by the relative motion of the source, screen and observer, which causes the diffraction pattern to move across the observer.
\subsubsection{The Fresnel scale}
\begin{figure}[ht]
\centering
\includegraphics[width=1.00\textwidth]{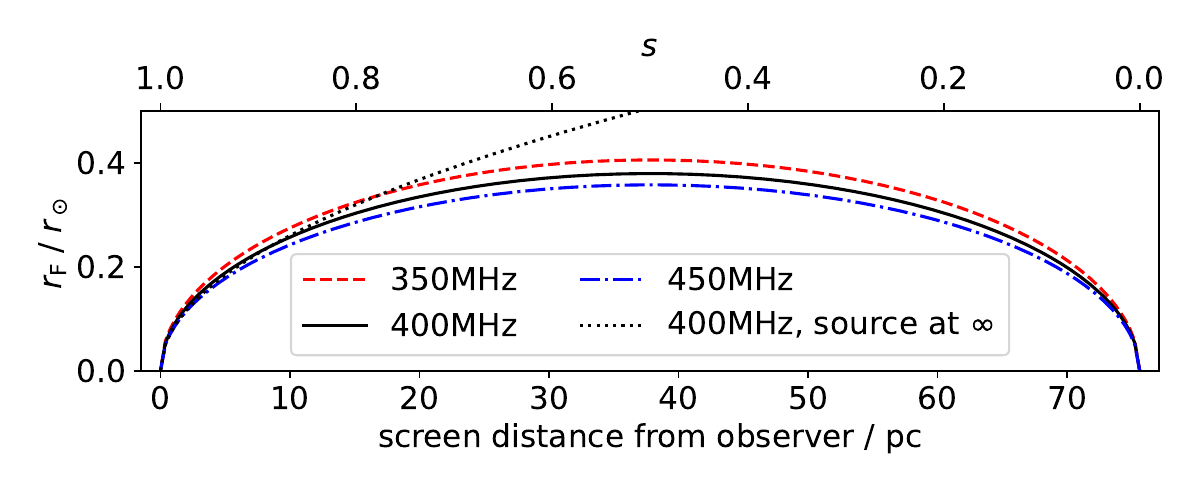}
\vspace{1ex}
\hspace{0.5em}
\includegraphics[width=0.95\textwidth]{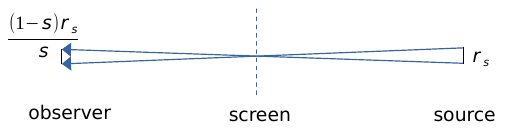}
\vspace{1ex}
\hspace{0.5em}
\includegraphics[width=0.95\textwidth]{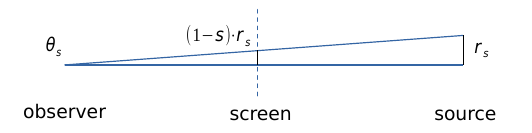}
\caption{
Top: Change in Fresnel scale with screen location and observing frequency for source at the distance of CU Vir. 
Middle: A small change in source location will shift the diffraction pattern in the aperture plane.
Bottom: Source structure as a convolution of the phase screen.}
\label{fig:r_f}
\end{figure}

\citet{1992RSPTA.341..151N} provides an excellent overview of scintillation physics.
The most important parameter for interpreting scintillation, the Fresnel scale, $r_\mathrm{F}$ is given by
\begin{equation}
    \label{eqn:r_f}
    r_F = \sqrt{\frac{\lambda d_e}{2\pi}} ,
\end{equation}
where $\lambda$ is the observing wavelength, and $d_e$ is the effective distance from the observer to the scattering screen.
Note that in contrast to \citet{1992RSPTA.341..151N} and much of the literature we do not make the simplifying assumption that the source is much further away from the observer than the scattering screen.
This means that we must take into account that the wavefront from the source at the screen is spherical rather than planar, and we do this by replacing the distance to the screen with the ``effective distance''.
This is given by \citet{2006ApJ...637..346C} as
\begin{equation}
    \label{eqn:d_e}
    d_e = s(1-s)d_\mathrm{source} ,
\end{equation}
where $d_\mathrm{source}$ is the distance from observer to source, and $s$ is the distance from source to screen as a fraction of distance from source to observer (i.e. 0 at the source, unity at the observer).
The top panel of \fig~\ref{fig:r_f} shows the change in $r_\mathrm{F}$ as a function of both screen location and frequency for a source at the distance of CU Vir.
$r_\mathrm{F}$ depends only weakly on observing frequency, and only weakly on screen location except for very close to observer or source.

\subsubsection{Scattering regimes}
\label{subsubsec:regimes}
Depending on the strength of scattering, weak or strong scintillation may be observed.
We follow \citet{2010ApJ...717.1206C} in using the Born variance, $m_b^2$ to quantify the scintillation strength.
$m_b^2$ is approximately equivalent to $\left(r_\mathrm{F}/r_\mathrm{diff}\right)^2$ using the notation of \citet{1992RSPTA.341..151N}, where $r_\mathrm{diff}$ is the physical distance on the screen over which the average phase variance is 1\,rad$^2$.
For the case of weak scintillation ($m_b^2\ll1$), the modulation index, $m$ (rms variation divided by mean brightness) is equal to $m_b$ and much less than unity, and the variations in intensity are highly correlated across observing frequency.
For weak scintillation, the characteristic spatial scale of the diffraction pattern is $r_\mathrm{F}$, and so the characteristic timescale is given by $r_\mathrm{F}/v$, where the velocity, $v$, is the relative motion of the screen, perpendicular to the line of sight. 

For the case of strong scintillation ($m_b^2\gg1$), where the medium introduces phase changes $\gg 1$\,radian over the first Fresnel zone, two distinct branches of scintillation emerge: diffractive scintillation, observable for sufficiently compact sources as large-amplitude ($m\sim$ unity), short-timescale ($t\sim r_\mathrm{F}/\left(m_b\cdot v\right)$), narrowband (fractional bandwidth $\sim1/m_b^2$) fluctuations.
Additionally, refractive scintillation is observable as slower($t\sim \left(r_\mathrm{F}\cdot m_b\right)/v$), weaker ($m\sim1/m_b$), broad-band variations due to mild focussing and defocussing effects \citep{1984A&A...134..390R}.

For scintillation in the transition region where $m_b^2 \sim 1$, the asymptotic solutions described in \citet{1992RSPTA.341..151N} for weak and strong scattering are not strictly applicable, however numerical simulations are useful to analyse scintillation in this regime \citep[][\sect~\ref{subsec:simulations}]{2010ApJ...717.1206C}. 

\subsubsection{Effects of source structure on scintillation}
Source size and structure are critical for determining any scintillation effects, which will not be observed for a sufficiently large source.
Here again we follow \citet{1992RSPTA.341..151N}, while noting that when the source is not in the far field, the relevant angular size of the source is that subtended at the screen, not the observer.

The effect of source structure can be understood by considering how the source will affect the 2D diffraction pattern at the location of the observer.
A small shift in the relative locations of the source, screen and observer will cause a shift in the location of this diffraction pattern (middle panel of \fig~\ref{fig:r_f}).
Source structure matters, because if the source is sufficiently large, then the maxima and minima of the diffraction pattern from different parts of the source will be washed out.
Consider a double source, with a linear separation $r_s$.
Each component will cast an almost identical diffraction pattern at the observer, but with an offset between them of $\left(1-s\right)r_s / s$ (\fig~1 middle panel).
The double source will therefore scintillate like a single point source, provided that the finest details in the scintillation pattern are large compared to this offset.
In the weak regime, this scale of the diffraction pattern is $r_F$. In the strong regime, the source size constraint for strong, diffractive scintillation is more stringent, and the scale is $r_\mathrm{F}/m_b^2$ (for refractive scintillation it is less stringent: $r_\mathrm{F}\cdot m_b^2$).
More generally, the observed diffraction pattern is the diffraction pattern of a point source convolved with the source brightness distribution.

Weak ISS is observed at gigahertz frequencies for sufficiently compact sources including AGN \citep[e.g.][]{2008ApJ...689..108L} and water megamasers \citep{2007ApJ...656..198V}. In case of scattering within tens of parsecs of Earth, this manifests as intra-hour variability \citep[e.g.][]{2020A&A...641L...4O}. At lower frequencies, slower variations are typically observed on timescales of weeks to months \citep{1986ApJ...307..564R}, corresponding to refractive scintillation in the strong scattering regime. While AGN are often too large to exhibit diffractive ISS, except where the scattering plasma is located very close to the observer \citep{2006A&A...446..185M, 2021MNRAS.502.3294W}, diffractive scintillation is universally observed for pulsars, manifesting as fluctuations in time and frequency in dynamic spectra \citep[e.g.][and references therein.]{2022ApJ...941...34S}

\subsection{Scattering towards CU Vir}
\label{subsec:ism}

Although CU Vir lies within the Local Bubble, defined by low neutral hydrogen density relative to the average ISM in the Milky Way \citep{2009SSRv..143..277S}, recent observations suggest that interstellar scattering ``screens'' are prevalent throughout this region. \cite{2025NatAs...9.1053R} analysed MeerKAT data on the bright millisecond pulsar J0437$-$4715 and discovered 21 distinct interstellar scattering screens through the presence of arcs in the secondary spectra. The authors place a limit on the interstellar thin screen number density of $\rho N \geq 0.14~{\rm pc}^{-1}$ on this line of sight, however due to selection biases limiting detection, they infer a higher screen number density $\rho N \approx 1~{\rm pc}^{-1}$. \cite{2022ApJ...927...99M} detect a scattering screen only 5.5\,pc away towards PSR~B1133+16. 
\citet{2021MNRAS.502.3294W} discovered a line of rapidly scintillating AGN inferred to be behind a narrow plasma filament at a distance of $\sim 4$\,pc, and make an order of magnitude estimate for the volume density of such filaments $\sim 10~{\rm pc}^{-3}$ - a few times larger than that of ordinary stars.
Thus, it is reasonable to expect that any given line of sight would likely intercept a number of interstellar scattering screens within a distance of 75\,pc, that could cause moderately strong scattering below $\sim 1$\,GHz.  

We note briefly that Galactic electron density models such as NE2001 \citep{2002astro.ph..7156C} and YMW16 \citep{2017ApJ...835...29Y} are not well-suited to estimating Galactic scattering on the line of sight to CU Vir. YMW16 predicts that a pulsar at the location of CU Vir would have a Dispersion Measure (DM) of 0.76\,pc\,cm$^{-3}$. However, the lowest DM pulsar used to determine the relationship between DM and scattering used by YMW16 has a DM of 3\,pc\,cm$^{-3}$ \citep{2015ApJ...804...23K,1986ApJ...311..183C}. For a general discussion of the difficulties in building a model of Galactic scattering, the reader is referred to \citet{2017ApJ...835...29Y}, \sect~1.
\subsection{Salient features of Electron Cyclotron Maser Emission}
\label{subsec:ecme_properties}

Electron Cyclotron Maser Emission is one of the two coherent mechanisms behind the production of intense (high brightness temperature) radio bursts in main-sequence stars, sub-stellar objects and planets; the other mechanism is plasma emission \citep[e.g.][]{melrose2017}.
ECME is produced by a non-thermal electron population with an unstable energy distribution in presence of a magnetic field. ECME is favoured when the local plasma frequency is less than the electron gyrofrequency, a situation that is easily achieved in the magnetospheres of MRPs thanks to their kG-strength surface magnetic fields \citep[e.g.][]{trigilio2008}.

The key properties of ECME include very high brightness temperature \citep[$\gtrsim 10^{24}$\,K have been observed,][]{treumann2006}, very high circular polarization \citep[$\sim 100\%$, e.g.][]{trigilio2000} and very high directivity \citep[$\lesssim 1^\circ$ for an individual pulse of radiation,][]{dulk1985}. Note that the fact that the emission is highly polarised has no effect on the scintillation \citep{2001Ap&SS.278..135M}. It is an intrinsically narrowband emission with the frequency of emission ($\nu$) being proportional to the local electron gyrofrequency $\nu_\mathrm{B}\approx2.8B$ \citep[$\nu_\mathrm{B}$ is in MHz and $B$ is in Gauss; e.g.][]{dulk1985}. As a result, higher frequencies are produced closer to the star where the magnetic field is stronger and vice-versa. 

\begin{figure}
    \centering
    \includegraphics[trim={3cm 3cm 0 2cm},clip,width=0.9\linewidth]{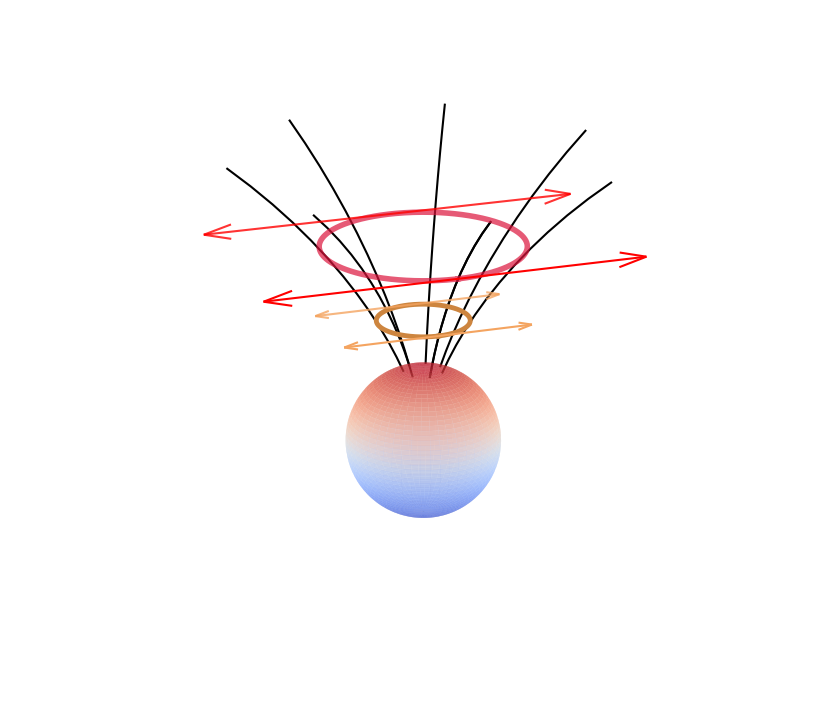}
    \caption{Illustration of the tangent plane beaming model of ECME \citep{trigilio2011} from a star with an axisymmetric dipolar magnetic field. A set of example magnetic field lines coming out of one of the magnetic hemispheres are shown in black solid lines. The two auroral rings act as emission sites for ECME at two different frequencies with the one closer to the star emitting at a higher frequency. According to the tangent plane beaming model, the emission is directed tangential to the auroral rings. The arrows illustrate directions of ECME produced in a given set of points on the two auroral rings. Note that each point on the auroral rings acts as sites of production of ECME, though the observer `sees' only a small fraction of the rings for which the ECME beams align with the line of sight as the star rotates.}
    \label{fig:ecme_beaming_model}
\end{figure}

In the case of MRPs, the magnetic fields can typically be approximated as axisymmetric dipoles inclined to the stellar rotation axes \citep[e.g.][]{shultz2018}. In that case, the regions producing ECME at a given frequency are the regions with the same magnetic field, which lie on a ring (called the ``auroral ring'') around the magnetic axis (\fig~\ref{fig:ecme_beaming_model}).
Assuming that ECME is produced in the fundamental harmonic along a magnetic field line of equatorial radius $L$ (in units of stellar radius $R_*$), the radius of the auroral ring can be approximated (when $L>>1$) as $\mathcal{R}\approx \sqrt{B_\mathrm{p}/(BL)}$\footnote{This approximate expression makes the assumption that the emission sites are sufficiently close to the magnetic dipole axis so that the $\theta$-component of the magnetic field on the auroral circle can be neglected as compared to the radial component. Without this assumption, the radius of the auroral circle needs to be solved numerically \citep[to solve \eqn~1 of ][]{das2020a}.}, where $B_\mathrm{p}$ is the polar magnetic field strength and $B$ is the magnetic field strength along the auroral ring under consideration. Using $\nu=2.8B$ (in MHz, in the fundamental harmonic), the approximate diameter of the auroral ring producing ECME at frequency $\nu$ is given by
\begin{align}
    2\mathcal{R}&\approx2\sqrt{\frac{2.8B_\mathrm{p}}{\nu L}}\label{eq:auroral_ring_size}
\end{align}
Using $B_\mathrm{p}= 4000$ G, $\nu=400$ MHz and $L=20$\footnote{Justification for this value: The sites of particle acceleration is believed to lie at the outermost parts of the stellar magnetosphere \citep[e.g.][]{leto2021}, thus $L\gtrsim R_\mathrm{A}$, where $R_\mathrm{A}$ is the Alfv\'en radius in unit of stellar radius. For CU\,Vir, $R_\mathrm{A}$ lies between 12--17 \citep{leto2006}.}, we obtain a diameter of $2.4\,R_*$ ($\approx 4.7\,R_\odot$ for CU\,Vir), thus larger than the star itself. However, as mentioned above, ECME is a highly directed emission so that not all the emission produced along the ring contributes to the observed emission. Based on the `tangent plane beaming model' \citep{trigilio2011}, the emission is directed tangential to the ring and perpendicular to the magnetic axis (see \fig~\ref{fig:ecme_beaming_model}). As a result, only a very small fraction of the auroral ring acts as the emission sites for the observed emission. For example, if we assume that the emission is beamed over a cone of width $5^\circ$ about the direction tangent to the auroral ring, the emission sites will consist of two segments lying diagonally opposite to each other on the relevant auroral ring, each of projected width (on a plane perpendicular to the line of sight) $\approx \mathcal{R}\theta^2/2$, where $\theta=5^\circ$. For the values considered above, this equals $\approx 0.0045\,R_*$ ($\approx 0.01\,R_\odot$ for CU\,Vir), which is much smaller than the size of the star itself. Thus, the combination of high directivity and narrow bandwidth of ECME is what makes the effective emission sites extremely compact, thus making it necessary to consider phenomena such as scintillation.

The compact nature of the emission sites is also supported by the high brightness temperature of ECME. CU\,Vir has been observed to produce a flux density in excess of 10 mJy over $400$ MHz to $>1$ GHz. Using a distance of $75.6$ pc and source size of $0.01\,R_\odot$, we infer a brightness temperature of $\sim 10^{17}$ K, which is consistent with brightness temperature expected for ECME.

\begin{figure}
    \centering
    \includegraphics[trim={2.5cm 0 0 0},clip,width=0.9\linewidth]{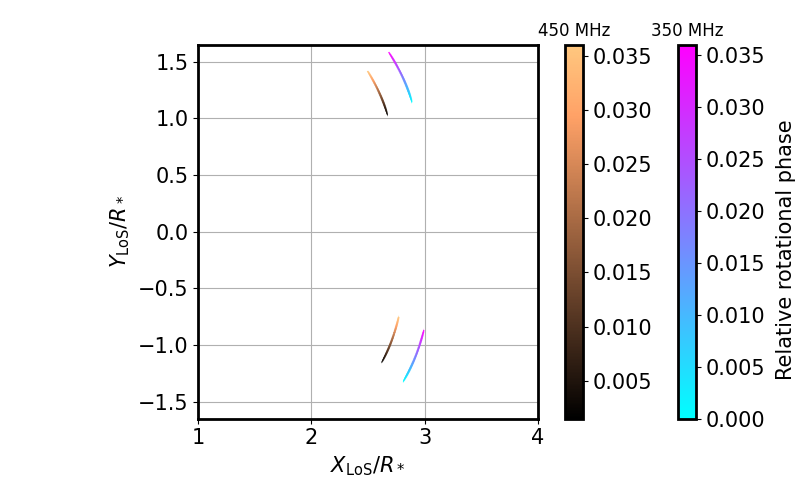}
    \caption{The motion of the emission sites producing ECME at two frequencies for a star like CU\,Vir (polar magnetic field strength of 4\,kG, inclination angle of $46.5^\circ$ and obliquity of $87^\circ$) over $\approx 0.04$ rotation cycle ($\approx 30$ minutes). We have used the co-ordinate systems used in \citet{das2020a} (see their Appendix B).}
    \label{fig:source_motion}
\end{figure}

While the emission sites at a given rotational phase and frequency are extremely compact, they can move significantly on the sky plane as a function of frequency, and as the star rotates.
To estimate the displacement of the emission sites on the sky plane, we use the 3D framework of \citet{das2020a}. Assuming that ECME is produced in the ordinary (O-mode) at the fundamental harmonic along magnetic field loops with equatorial radii of $20\,R_*$ and using the stellar magnetospheric parameters of CU\,Vir listed in \sect~\ref{sec:introduction}, we simulated the ray paths 
assuming a simplistic $1/r$ density profile\footnote{I.e. $n_\mathrm{p}=n_\mathrm{p0}/r$ \citep{leto2006}, where $n_\mathrm{p}$ is the electron number density at a radial distance of $r$ in units of stellar radius. We set $n_\mathrm{p0}=10^8\,\mathrm{cm^{-3}}$. This value corresponds to the minimum value of this parameter considered by \citet{leto2006}. The reason for choosing a smaller value is that more advanced magnetospheric model shows that the thermal plasma is mostly confined in thin discs \citep[e.g.][]{townsend2005,berry2022}. Refraction of emission due to this profile is included in the model. We assume that since ECME at $400$ MHz arises significantly above the magnetic poles, the radiation will not experience the high density region.}. This exercise shows that at a given rotational phase, indeed the emission sites consist of two extremely compact regions (projected on the sky-plane) with sizes much smaller than the size of the star, though they can move significantly as the star rotates, and change location as a function of frequency.
This is illustrated in \fig~\ref{fig:source_motion}.

\subsection{Scintillation of ECME emission}
\label{subsec:scint_ecme}
In \sect~\ref{subsec:scintillation} we described the criteria required for scintillation to take place;
in \sect~\ref{subsec:ism} we provided evidence of the ubiquity of scattering that would cause strong scintillations from CU Vir.
Finally, in \sect~\ref{subsec:ecme_properties} we sketched out the features of ECME which will determine whether scintillation can be observed.
We are now ready to assess whether ECME emission, according to our model, could display strong scintillations provided that there is sufficient scattering along the line of sight.

The emission comes from two emission regions separated by several solar radii.
Let us first consider each emission region individually, and for the monochromatic case.
In this case, each emission region is a small patch $\sim0.01\,R_\odot$ in extent.
This is an order of magnitude smaller than $r_\mathrm{F}$ for all screen distances except for those extremely close to either the observer or the star (top of \fig~\ref{fig:r_f}).
The finite source size certainly would act to suppress scintillation due to ionized media within $\sim$4\,pc of the star; however other propagation effects due to this plasma would still be possible.
If these effects caused sufficient angular broadening of the emission, then this could act to suppress scintillation due to a more distant scattering screen.
In the absence of these effects, however, for a scattering at an intermediate location, finite source size would only strongly affect scintillation for screens with scattering strength $m_b^2\gtrsim100$.

Next, we consider the effect of having two point sources separated by approximately 7 solar radii ($\sim$20\,$r_\mathrm{F}$).
This is several times the Fresnel scale for any screen geometry, and so the scintillation pattern from each will be uncorrelated.
Therefore, the observed scintillation pattern will simply be the incoherent sum of two approximately equal contributions.
For diffractive scintillation, this will reduce the modulation index from 1 to $1/\sqrt{2}$, but the timescale and scintillation bandwidth will be unaffected.
Refractive scintillation is caused by structures of spatial scale $r_\mathrm{F}\cdot m_b$, so the assumption of independent scintillation may break down for very long timescales.

Next we consider the effects of the motion of the emission region (1$R_\odot$ in 30 minutes $\approx 400$\,km\,s$^{-1}$) as the star rotates.
This is large compared to typical velocities inferred from ISS, which are typically of order 50\,km\,s$^{-1}$, and also large compared to the Earth's orbital motion $v_\oplus\approx30$\,km\,s$^{-1}$, so except for scattering very close to the Earth, the motion of the emission region will likely dominate.
 
Finally, we consider the fact that the location of the source region changes with frequency.
The shift of $\sim 0.5R_\odot$ across the observing bandwidth is comparable to the Fresnel scale, and so is likely to have a significant effect, in terms of shifting the location of the scintillation pattern on the ground.
This introduces changes across the frequency band which are very similar to those normally seen with time; for instance, variability across the frequency band could also be seen for weak or refractive scintillation, with a modulation index $<$1.
For diffractive scintillation therefore, the shift in source location with frequency may cause the scintillation pattern to change more rapidly with frequency than it otherwise would. 

\subsection{Simulating Scintillation of CU Vir}
\label{subsec:simulations}

In the foregoing sections we have described the various regimes of scintillation in terms of the scattering strength, distance to screen and source, etc., relying predominantly on \citet{1992RSPTA.341..151N}.
We can also use the \textsc{scintools} python package \citep{2010ApJ...717.1206C,2020ApJ...904..104R} to produce a random realization of a dynamic spectrum for a set of observing and scintillation parameters.
Below we list the parameters that we must set, and justify them where appropriate.

A reasonable starting point for the scattering strength is $m_b^2=9$ on the basis that this is the reciprocal of the fractional bandwidth (measured approximately by visual inspection) of the spectral features seen by \citet{2021ApJ...921....9D} (\fig~11; see also our \fig~\ref{fig:dynspec}).
We note that \citet{2010ApJ...717.1206C} caution against the use of the scintillation bandwidth for estimating scintillation strength; showing, through the use of simulations, that the measured fractional bandwidth can vary over an order of magnitude even when the underlying scintillation strength is constant.
We further note that the change in emission region with frequency for ECME may make this an overestimate of the scattering strength.
For this reason, we reduce the scattering strength by a factor of 2 to $m_b^2=4.5$.

We note that the descriptions of the ``strong'' and ``weak'' regimes given in \sect~\ref{subsubsec:regimes} are asymptotic, for $m_b^2\gg1$ and $m_b^2\ll1$ respectively, and so this scattering strength is potentially in an intermediate regime \citep{2006ApJ...636..510G}.
However, as we discuss in the next section the result is a dynamic spectrum that has the qualitative features of strong-regime scintillation.

We assume that the screen is located halfway between source and observer, and so this sets $r_\mathrm{F}$ to be 0.38$R_\odot$ at the central frequency\footnote{When running the code we find it more convenient to set $r_\mathrm{F}$ to unity, and give all other parameters in terms of $r_\mathrm{F}$}.

The simulation package does not require us to set a screen velocity, but we adopt 400\,km\,s$^{-1}$ (the source motion calculated in \sect~\ref{subsec:scint_ecme}) to provide a mapping between units of $r_F$ and time on the x-axis.
We set the power law index of the phase screen to $\alpha=5/3$, the value for Kolmogorov turbulence.
Anisotropic phase screens can be specified, but we choose an isotropic one.

We may also specify an inner scale for the turbulence, for which $l_0=100$\,km($<10^{-3}r_\mathrm{F}$) would be appropriate \citep{1990ApJ...353L..29S,2009MNRAS.395.1391R}.
However, the effect of an inner scale and the effect of finite source size are very similar: both are a 2D convolution of the 2D phase screen with a small kernel that smooths over the smallest structures.
Source structure as a convolution of the \emph{screen} rather than the diffraction pattern can be understood from the point of view of the observer.
The finite source size is then a convolution of the screen with the source structure scaled by $(1-s)$ (\fig~1, lower panel).
Therefore, we specify an inner scale of 0.012$r_\mathrm{F}$, which is equivalent to smoothing with a circular Gaussian of this width.

We set the resolution of the phase screen to 0.014$r_\mathrm{F}$ and its dimensions to $x, y = 1000, 500$ pixels.
Given our assumed velocity, this makes the duration of the simulated spectrum almost exactly 0.2 rotational periods ($\approx2.5$hr).
We match the bandwidth of our simulation to the data, but with a higher frequency resolution of 100 channels.

As discussed in the preceding section, the expected observed scintillation pattern for ECME emission from CU Vir is the sum of two independent scintillation patterns. Therefore, we run our simulation  (with the same parameters, but producing different random realizations of the phase screen), producing two dynamic spectra, then sum the results and divide by two. 

\section{RESULTS}
\label{sec:results}
\begin{figure*}[ht]
\centering
\includegraphics[width=1.0\textwidth]{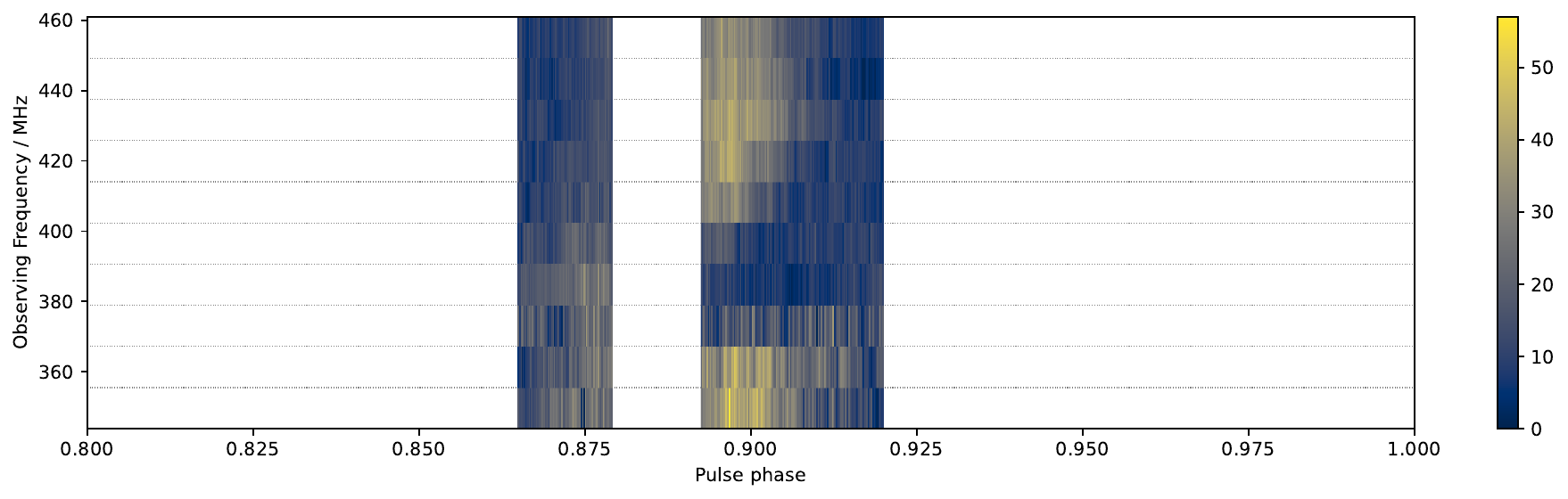}
\includegraphics[width=1.0\textwidth]{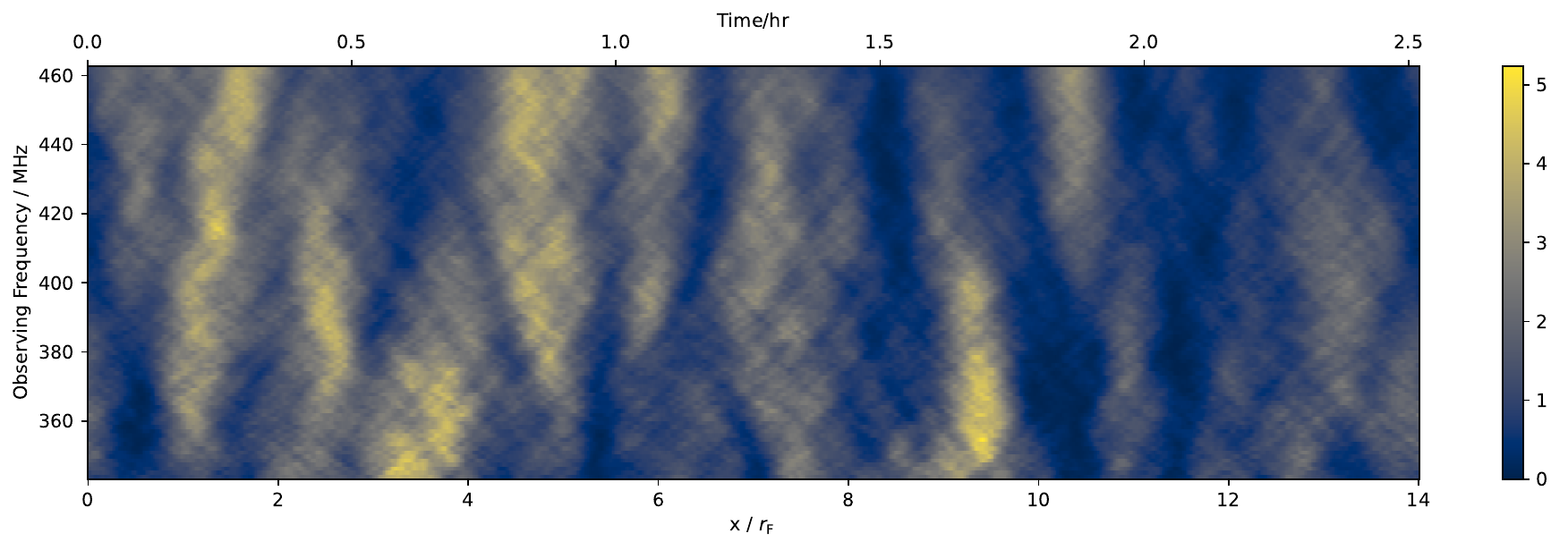}
\caption{
Top: Observed dynamic spectrum of a burst from CU Vir. This is precisely the same data shown in \fig~11 of \citet{2021ApJ...921....9D}. The colour scale has units of mJy and is linear with a range of 0 to the maximum of the data. Dotted lines indicated channel boundaries. Only the timerange corresponding to the published figure is shown.
Bottom: A simulation of strong scintillation with the parameters described. The colour scale is similar to the top figure in that it is linear and covers zero to the maximum of the data, but the units are relative to the known mean flux. Time and frequency axes are matched to the top panel, but the model is of a stochastic process and so there is no expectation that features will match in time. $r_\mathrm{F}$ is for the central frequency. A wider span is simulated than is shown in the top panel to give a more extensive picture.}
\label{fig:dynspec}
\end{figure*}

We compare our simulated dynamic spectrum with observations in \fig~\ref{fig:dynspec}.
The top panel shows precisely the same data as the right panel of \fig~11 of \citet{2021ApJ...921....9D}.
A variability between neighbouring points in time for each spectral channel is evident in these data; however, by comparison with off-pulse data (not shown) we have satisfied ourselves that this is mostly just (white) instrumental noise.

Both panels of \fig~\ref{fig:dynspec} have matching x axes, and so the three x-scales (pulse phase, time in hours, and a spatial scale in terms of $r_\mathrm{F}$) are shared between them.
The $r_\mathrm{F}$ scale is the most fundamental, and is linked to the timescale by our assumed velocity.
This is then readily converted into pulse phase using the rotational period of the star.
We reiterate that with the lower panel we are simulating a stochastic process, and so the absolute values of the scales are arbitrary: there is no expectation that particular features will line up between the data and the model.
The reason for providing a full 2.5 hours of simulated data is to provide the reader with sufficient simulated data to be able to see the kinds of behaviours exhibited by scintillation of this strength.

Diffractive scintillation from pulsars varies considerably depending on the line of sight, and the reader is referred to \citet{2006ApJ...637..346C} to see dynamic spectra for a range of scintillation regimes and velocities, for both real data and simulated data.
As noted in the previous section, the chosen scintillation strength ($m_b^2=4.5$) places us in an intermediate regime between weak and strong scintillation.
However, the dynamic spectrum in the bottom panel of \fig~\ref{fig:dynspec} contains many of the expected features of scintillation in the strong regime.
The main features that can be seen are scintles which have a width of a fraction of $r_F$, and a fractional bandwidth of $\sim$20\% (i.e. spanning half or more of the simulated bandwidth).
Broadly speaking these numbers are in line with the approximate relations given in \sect~\ref{subsubsec:regimes} \citep[see also][]{1992RSPTA.341..151N}.
There is a weak, fine ``crisscross'' pattern which is also seen when pulsars scintillate \citep{2001ApJ...549L..97S}; however, if it is present in the data presented in the top panel, the spectral resolution is not sufficient to resolve it.
Refractive scintillation would be expected to be observed in this regime, and should be somewhat weaker than the diffractive scintillation, with a scale of a few $r_F$.
The relative lack of intensity in the last quarter of the figure is consistent with this timescale, but a longer simulation would be required to clearly diagnose the presence of this effect.

As discussed in \sect~\ref{subsec:simulations}, there are two important effects that are non-trivial for us to model.
However, the $r_\mathrm{F}$ scale (at the bottom of the lower panel of \fig~\ref{fig:dynspec}) is useful in providing an understanding these two effects.
First, we have assumed that the scintillation patterns from the two emission points combine incoherently.
This is not true for the large-scale structures in the phase screen which cause refractive scintillation.
However, since the two emission points are approximately 20$r_\mathrm{F}$ apart, this will not be relevant for the temporal scales considered here.
It may be relevant when considering pulse-to-pulse variations.

Secondly, and more importantly, the changing location of the emission with frequency could not easily be simulated.
Therefore, the change in brightness across the band seen in the simulations is not due \emph{only} to the fact that the scintillation is moderately strong.
The change in location is approximately 1$r_\mathrm{F}$ across the observing bandwidth.
We can therefore quantify the unmodelled change in brightness across the band due to source motion, even though it has not been simulated.
The \emph{additional} change in brightness with frequency induced by this change in frequency, is therefore equivalent to a motion of 1$r_\mathrm{F}$ in the $x$ direction.
It can be seen that the separation of scintle maxima in the $x$ direction is typically a little more than 1$r_F$.
Therefore, the effect of source motion will be to reduce the size of the scintles along the frequency axis, so that instantaneously (i.e. at a particular location along the $x$-axis) there are approximately two rather than approximately one.
We note that the movements of the emission region with frequency and with time are roughly orthogonal (see \fig~\ref{fig:source_motion}), and so frequency and time variations should be uncorrelated provided that the motion of the source region dominates over the motion of the Earth and screen.
The dynamic spectrum as simulated does show a moderate drift in frequency for some scintles (i.e. they are not perfectly vertical), however there is no obvious preference to the direction of drift. 

The simulation and the data are remarkably similar in that the width in time of the emission seen from CU Vir in the low-frequency radio observations is similar to the width of the scintles in both frequency and time.
This is partially by construction, since we have chosen our simulation parameters on the basis of the approximate bandwidth of the features seen in the data.
Nonetheless, the simulation shows that the bandwidth-limited bursts seen by \citet{2021ApJ...921....9D} can plausibly be explained by diffractive ISS.

\section{DISCUSSION}
\label{sec:conclusions}
While we believe the simulation shows that the observed bandwidth limited features can be explained by diffractive ISS, this is not the only explanation.
\citet{2021ApJ...921....9D} speculated that it could be due to the superposition of two unrelated bursts occurring at the same time, or due to propagation effects within the magnetosphere.
The trapped plasma in the stellar magnetosphere causes different frequencies to deviate by different extents \citep{trigilio2011}. Thus, even though the intrinsic ray directions are confined to planes parallel to each other (at all frequencies) in the tangent plane beaming model, after passing through the magnetosphere, they no longer remain so. The result is that the beams at different frequencies cross the line of sight at different times as the star rotates, creating a sweep on the dynamic spectrum. 
This drift caused by magnetospheric propagation effect and stellar rotation, could also provide an alternative explanation behind the observed spectral feature. In that scenario, the enhancements above and below 390 MHz observed over phases 0.89-0.92 represent two separate components of ECME, both of which are drifting to higher frequencies with time, and the apparent dip in flux density around 390 MHz is caused by the slight difference in their arrival times. 

Additionally, unless these processes radically alter the size of the emission site (as viewed by any scattering screen) there is no reason why multiple effects might not be present.
Therefore, in order to correctly interpret such observations in the future, it may be necessary to disentangle several propagation effects.
Below we offer some suggestions of how further observations might eliminate or better quantify the impact of ISS. 

The stochastic nature of scintillation is one aspect that is falsifiable.
If the null at approximately 390\,MHz persists over multiple observations, especially ones widely separated in time, this would almost certainly be inconsistent with interstellar scattering (narrowband spectral features can persist for longer periods \citep{1997MNRAS.287..739R}, but these are due to larger structures).

Our simulated dynamic spectrum shows only modest drifts of the frequency of emission with time, with no clear preference in the direction of drift.
Much faster drifts than in frequency than seen in our simulations, or those which are preferentially in one direction, may not be attributable to scintillation effects, although full simulation of the moving source with frequency, with a full range of screen distances, velocities, and turbulence anisotropies, would be required in order to be conclusive.

On the other hand, ISS can be definitively diagnosed with simultaneous observations of the star, in the same frequency range, using instruments separated by $\sim1000$\,km or more.
This is sufficient for each instrument to sample a different region of the diffraction pattern, ensuring that the two instruments will observe approximately the same temporal variation with a measurable time delay \citep{2000aprs.conf..147J}.
Naturally, observing different behaviour at different observing sites rules out intrinsic variability.
At the time of writing, the SKA-Low is rapidly being deployed, and once it reaches sufficient sensitivity it would be possible to observe with both the GMRT and the SKA-Low over the relevant frequency range.

Another consequence of strong scatter at sub-GHz frequencies is that there will still be significant scintillation at higher frequencies. 
For the scattering strength assumed in this work, the transition frequency between strong and weak scatter \citep[see e.g.][]{1998MNRAS.294..307W} would be 870\,MHz; and the modulation index at 2\,GHz would still be $\sim$30\%.
If pulse-to-pulse variations at higher frequencies were found to be significantly smaller, this would be a strong argument against ISS.
With sufficient data, it may even be possible to diagnose ISS if the modulation index varies as a function of frequency according to the expected $\lambda^{17/12}$ relationship.
However, until it is ruled out, ISS also provides a plausible mechanism for pulse-to-pulse variation.

With the advent of the SKA era, it will be possible to study CU Vir with unprecedented sensitivity over a wide frequency range.
It will also become possible to study many stars already known to emit in the radio \citep{2024PASA...41...84D} at high sensitivity for the first time.
Dynamic spectra of strong ISS encode information about the size of the emitting region, and motion of the emitting region with time and frequency.
Determining whether and how this information can be gleaned from stellar dynamic spectra, especially since there will be intrinsic brightness variations such as the pulses themselves, is beyond the scope of this paper.
However, the combination of wide-bandwidth observations, VLBI observations of scintillating sources \citep{2010ApJ...708..232B} and measurements of annual changes \citep[e.g.][]{2021MNRAS.502.3294W}, provide a formidable range of information.
Therefore, it is possible that with extensive observations, scintillometry will become a powerful tool for testing coherent stellar emission mechanisms.

\begin{acknowledgement}
This work has made use of data from the European Space Agency (ESA) mission
{\it Gaia} (\url{https://www.cosmos.esa.int/gaia}), processed by the {\it Gaia}
Data Processing and Analysis Consortium (DPAC,
\url{https://www.cosmos.esa.int/web/gaia/dpac/consortium}). Funding for the DPAC
has been provided by national institutions, in particular the institutions
participating in the {\it Gaia} Multilateral Agreement.
\end{acknowledgement}

\bibliography{refs}


\end{document}